\documentclass[3p,times,procedia]{elsarticle}
\usepackage{nupha_ecrc}
\usepackage{hyperref}
\hypersetup{
    colorlinks=true,
    linkcolor=blue,
    filecolor=magenta,      
    urlcolor=cyan,
}

\volume{00}

\firstpage{1}

\journalname{Nuclear Physics A}

\runauth{B.~Duclou\'e et al.}

\jid{nupha}

\jnltitlelogo{Nuclear Physics A}

\usepackage{amssymb}

 \usepackage{bm,amsthm}
 
 \usepackage[mathscr]{eucal}
\usepackage{color}
\usepackage{caption}
\usepackage[normalem]{ulem}
\usepackage{setspace}
\usepackage{multirow}
\usepackage{graphicx}

\usepackage[figuresright]{rotating}

\long\def\comment#1{ }
\newcommand{\abar}{\bar{\alpha}_s}
\newcommand{\eqn}[1]{Eq.~(\ref{#1})}
\newcommand{\eqref}[1]{~(\ref{#1})}
\newcommand{\beq}{\begin{eqnarray}}
\newcommand{\eeq}{\end{eqnarray}}

\newcommand{\dif}{{\rm d}}
\newcommand{\rmd}{{\rm d}}
\newcommand{\rme}{{\rm e}}

\newcommand{\del}{\partial}

\newcommand{\minus}{\!-\!}
\newcommand{\order}[1]{\mathcal{O}{(#1)}}
\newcommand{\xbj}{x_{\rm\scriptscriptstyle Bj}}

\newcommand{\kt}{k_\perp} 
\newcommand{\bq}{\bm{q}}

\newcommand{\bx}{\bm{x}}
\newcommand{\by}{\bm{y}}

\newcommand{\bz}{\bm{z}}

\begin{document}

\begin{frontmatter}



\dochead{XXVIIIth International Conference on Ultrarelativistic Nucleus-Nucleus Collisions\\ (Quark Matter 2019)}

\title{Collinear resummations for the non-linear evolution in QCD at high energy}



\author[sac,edb]{B.~Duclou\'e}
\ead{bertrand.ducloue@ed.ac.uk}

\author[sac]{E.~Iancu}
\ead{edmond.iancu@ipht.fr}


\author[col]{A.H.~Mueller}
\ead{ahm4@columbia.edu}

\author[sac]{G.~Soyez}
\ead{gregory.soyez@ipht.fr}

\author[ect]{D.N.~Triantafyllopoulos}
\ead{trianta@ectstar.eu}

\address[sac]{Universit\'{e}  Paris-Saclay, CNRS, CEA, Institut de physique  th\'{e}orique, 91191, Gif-sur-Yvette, France}

\address[edb]{Higgs Centre for Theoretical Physics, University of Edinburgh, Peter
Guthrie Tait Road, Edinburgh EH9 3FD, UK}

\address[col]{Department of Physics, Columbia University, New York, NY 10027, USA}

\address[ect]{European Centre for Theoretical Studies in Nuclear Physics and Related Areas (ECT*)\\and Fondazione Bruno Kessler, Strada delle Tabarelle 286, I-38123 Villazzano (TN), Italy}

\begin{abstract}
When computed to next-to-leading order in perturbative QCD, the non-linear Balitsky-Kovchegov (BK)
equation for the high-energy evolution of the dipole-hadron scattering appears to be unstable. We show that
this instability can be avoided by using the rapidity of the dense hadronic target (instead of that of the dilute
dipole projectile) as the evolution time. Using this variable, we construct a collinearly-improved version
of the BK equation, where the dominant radiative corrections to the kernel --- those enhanced by
double collinear logarithms --- are resummed to all orders.

  \end{abstract}

\comment{
}

\end{frontmatter}



\bigskip
\paragraph{Introduction: High-energy evolution and gluon saturation} We would like  to study the high-energy evolution of a hadron (proton or nucleus)
wavefunction in the vicinity of saturation, i.e. in the high parton density regime where gluon
occupation numbers are of order $1/\alpha_s$ and non-linear effects like gluon recombination
play an important role. This evolution proceeds via the successive emission of soft gluons with smaller
and smaller longitudinal momentum fraction $x=k^-/P^-$; here, $k^-$ is the (light-cone) longitudinal
momentum of the emitted gluon and $P^-$ is the respective momentum of the parent nucleon,
assumed to be an energetic left mover. The
phase-space for this evolution is controlled by the rapidity difference $\eta\equiv \ln(P^-/k^-)=\ln(1/x)$.
The probability for one soft gluon emission is of order $\abar\eta$, with $\abar\equiv \alpha_sN_c/\pi$.
When $\abar\eta\gtrsim 1$, gluon emissions must be iterated, which one can do by solving a
evolution equation in $\eta$, that we intend to construct. With decreasing $x$,
or increasing $\eta$, the gluon density grows, hence saturation occurs for smaller and smaller
gluon transverse sizes $\Delta x_\perp$, or, equivalently, for larger and larger transverse momenta
$\kt\sim 1/\Delta x_\perp$. The typical transverse momentum for which a gluon with energy fraction
$x$ reaches saturation is known as the {\it saturation momentum} $Q_s(\eta)$. The previous
discussion shows that this scale is increasing with $\eta$. For sufficiently large 
$\eta$ at least, this increase is exponential: $Q_s^2(\eta)\simeq Q_0^2\rme^{\lambda_s \eta}$,
where the {\it saturation exponent} $\lambda_s$ is a number of $\order{\abar}$,
to be obtained from solutions to the aforementioned evolution equation. 

\paragraph{Deep inelastic scattering at small Bjorken $x$}
To study this evolution in a gauge-invariant manner, we consider the scattering between the hadron and a small projectile, with transverse resolution 
of the order of the saturation scale. Two important examples are deep inelastic scattering (DIS)
at small Bjorken $\xbj$ and particle production in proton-nucleus ($pA$) collisions. For definiteness, here
we shall mostly refer to DIS. The Bjorken $\xbj$ variable is defined as
$\xbj\equiv \frac{Q^2}{2P\cdot q}=\frac{Q^2}{2P^- q^+}$,
where the second equality holds in a frame where the
virtual photon $\gamma^*$ is an energetic right-mover, 
with  4-momentum $q^\mu\equiv (q^+, q^-, \bq_\perp)= (q^+, -Q^2/2q^+,\bm{0}_\perp)$, whereas the proton is a left-mover with $P^\mu=\delta^{\mu-}P^-$.
At high-energy or small Bjorken $\xbj$, this process admits
a {\it dipole factorization}: $\gamma^*$ fluctuates into a quark-antiquark
pair in a colour-singlet state (the ``dipole''), which then inelastically scatters off the proton target.
Indeed the condition $\xbj\ll 1$ implies that the  lifetime $\Delta x^+\simeq 2q^+/Q^2$ of the $q\bar q$
fluctuation is much larger than the longitudinal extent $1/P^-$ of the target.  

The dipole-proton interaction is controlled by the target gluons with longitudinal momenta $k^-\gtrsim Q^2/2q^+$,
which fully  overlap in time with the $q\bar q$
fluctuation  ($1/k^-\lesssim \Delta x^+$).
The fraction $x=k^-/P^-$ of the softest such gluons
coincides with $\xbj$, so the latter fixes
the rapidity phase-space $\eta=\ln(1/x)$ for the target evolution.
From now on we shall identify these two variables, $\xbj\equiv x$, and use the
simpler notation $x$ for both.

\paragraph{Target vs. dipole evolution} Thus, the physical problem that we are interested in --- the evolution of the proton wavefunction with decreasing $x$ --- is tantamount to computing
the $x$-dependence of the DIS structure functions  
for fixed $Q^2$ (eventually chosen of the order of
$Q_s^2(\eta)$,  to study gluon saturation in the target). However, this
problem is complicated,
due to the need to compute gluon emissions in a dense environment,  and 
so far it has been worked out only  to leading order (LO) accuracy, leading to the JIMWLK equation \cite{Gelis:2010nm}.

A simpler approach is to follow the evolution of the 
dipole. Soft gluon emissions from the dipole
occur like in the vacuum and non-linear effects exclusively refer to multiple scattering.
This viewpoint leads to the Balitsky hierarchy (in particular, to the BK equation), 
which is currently known to  next-to-leading-order  (NLO) 
accuracy~\cite{Balitsky:2008zza,Balitsky:2013fea,Kovner:2014lca,Lublinsky:2016meo}.  
Our purpose is to go beyond LO, so we
shall adopt the dipole evolution in what follows. But this comes with a drawback: 
the NLO version of the BK equation turns out to be {\it unstable} \cite{Lappi:2015fma}.

This instability can be traced back to two facts: \texttt{(i)} The rapidity  difference $ Y\equiv \ln({q^+}/{q_0^+})$ between the projectile and the softest gluons in the dipole wavefunction is different from the rapidity $\eta=\ln(1/x)$ of the target. \texttt{(ii)} The LO evolution 
 in $Y$ can violate the correct {\it time ordering} of the successive gluon emissions.
 
Concerning point  \texttt{(i)}, we observe that the softest gluons from the dipole wavefunction
which matter for the scattering are those with a 
transverse momentum $\sim Q_0$ and a
lifetime ${2q_0^+}/{Q_0^2}\sim 1/P^-$; this implies
\beq
    Y=\,\ln\frac{q^+}{q_0^+}=\,\ln\frac{2q^+P^-}{Q_0^2}=\,\ln\frac{1}{x}+
\ln\frac{Q^2}{Q_0^2}\equiv \eta +\rho\,.
\label{Ydef}
\eeq
The difference  $Y-\eta=\rho\equiv\ln(Q^2/Q_0^2)$ is quite large for the interesting values
$Q^2\sim Q_s^2(\eta)$: the relative difference $(Y-\eta)/\eta\sim \lambda_s$
is of  $\order{\abar}$, so it formally matters at NLO and beyond. 

Concerning point  \texttt{(ii)}, we note that, in the LO evolution with $Y$, the successive
gluon emissions are explicitly ordered in longitudinal momenta, 
$q^+\gg k_1^+\gg k_2^+\gg\dots\gg q_0^+$, and  {\it implicitly} ordered in lifetimes
\beq\label{TOcond}
\frac{2q^+}{Q^2}\gg \frac{2k_1^+}{ k_{1\perp}^2}\gg  \frac{2k_2^+}{k_{2\perp}^2}\gg\dots\gg \frac{2q_0^+}{Q_0^2}\,,\eeq
since daughter gluons must live shorter than their parents. Although {\it implicitly assumed}, time-ordering is not {\it explicitly enforced} in the LO BK equation. In fact, it is violated by the {\it typical} emissions, whose transverse momenta obey $ Q^2\gg k_{1\perp}^2\gg k_{2\perp}^2 \gg\dots\gg Q_0^2$ and which
 give the dominant, double-logarithmic, contributions to the evolution of the dipole amplitude:
 a series of powers of $\abar Y\rho$. Such violations are unphysical and are corrected by
 higher-order perturbative corrections to the evolution in $Y$. Yet, the associated corrections
 are parametrically large and spoil the convergence of the weak coupling expansion.
  
 To understand that,
 notice that the effect of the time-ordering condition\eqref{TOcond} is to  
 reduce the rapidity phase-space available for the evolution from 
$Y$ to $Y-\rho\equiv \eta$. Accordingly, the correct double-logarithmic approximation
should be a series  in powers of $\abar(Y-\rho)\rho$, instead of the LO series in 
powers of $\abar Y\rho$. The difference between the two corresponds to a {\it tower} of series of radiative corrections enhanced by the double
 ``anti-collinear'' logarithm $\rho^2$: the dominant series includes all powers of
  $\abar\rho^2$, the subdominant one, those of $\abar^2\rho^2$, etc.
 In particular, the NLO BK equation includes the first (negative)
contribution proportional to $\abar\rho^2$~\cite{Balitsky:2008zza}
which makes the evolution {\it unstable} \cite{Lappi:2015fma} and
hence unsuitable for physical studies.

\paragraph{Collinear resummations in $Y$}The dominant series in  powers of $\abar\rho^2$ can be resummed to all orders
by enforcing time-ordering in the LO BK evolution \cite{Beuf:2014uia,Iancu:2015vea}. Two ``collinearly improved''
BK equations have been proposed \cite{Beuf:2014uia,Iancu:2015vea}, with seemingly promising results:
the ensuing equations are stable~\cite{Iancu:2015vea,Lappi:2016fmu}, they can be extended to full NLO
accuracy~\cite{Lappi:2016fmu} and they allow for good
fits to the HERA data for DIS at small $x$~\cite{Iancu:2015joa,Albacete:2015xza}.

A more recent study has revealed that these apparent successes
are deceiving~\cite{Ducloue:2019ezk}. The numerical studies
in~\cite{Lappi:2016fmu,Iancu:2015joa,Albacete:2015xza} have been
presented in terms of $Y$ instead of the physical
rapidity $\eta=Y-\rho=\ln(1/x)$. Also, in the DIS fits in~\cite{Iancu:2015joa,Albacete:2015xza}, 
the variable $Y$ has been abusively interpreted as $\ln(1/x)$.
When translating the respective results from $Y$ to $\eta$, 
 one finds \cite{Ducloue:2019ezk} an {\it
 unacceptably large resummation-scheme dependence}, that should be
 attributed to the uncontrolled,  subleading, double-logarithmic 
 corrections. 
 This large scheme dependence forbids any physical interpretation of
the results.  The problem is further complicated by the fact that the resummed 
BK evolution in $Y$ cannot be formulated as an initial-value problem
\cite{Ducloue:2019ezk}.
%

\paragraph{Recasting dipole evolution in terms of $\eta$}
These difficulties with the evolution in $Y$ 
can be avoided by using $\eta$ as an ``evolution time'' for the dipole projectile \cite{Ducloue:2019ezk}. 
This may not look very natural (recall that $\eta$ is the rapidity of the target), 
yet it can be unambiguously implemented in perturbation theory, 
via a change of variables $Y\to \eta\equiv Y-\rho$. And indeed,
Ref.~\cite{Ducloue:2019ezk} has used this trick to deduce the NLO BK
equation in $\eta$ from the respective equation in $Y$~\cite{Balitsky:2008zza}.
When the gluon emissions are ordered in $\eta$, they automatically
obey the correct time-ordering: indeed,  $\eta= \ln(P^-/k^-)$, hence ordering in $\eta$ is
tantamount to ordering with increasing $k^-$, or decreasing $\Delta x^+=1/k^-$. Besides, 
$\eta=\ln(1/x)$ is the right variable to be used in phenomenological studies 
of DIS. Last but not least, the evolution in $\eta$ is naturally formulated as 
an initial value problem.  

Since the proper time-ordering is now automatic, the evolution in $\eta$ is not affected by the large anti-collinear logarithms present in the evolution in $Y$. On the other hand, the ordering
in $k^+$ is not enforced anymore and it can be violated
by  {\it collinear} emissions, where the daughter gluon has a
much larger transverse momentum than the parent one.
Such collinear emissions are  {\em atypical} in the DIS context
(they don't matter for the double-logarithmic approximation), hence they are less problematic.
E.g., they yield double-logarithmic corrections to the BK
kernel (starting with NLO), but not to the solution to the BK equation  \cite{Ducloue:2019ezk}.
Besides, they are suppressed by saturation, which freezes the evolution for 
emissions with  $k_\perp\lesssim Q_s(\eta)$. 
This being said, the effects of the double collinear logarithms in the kernel are
still quite large and moreover they accumulate during the evolution. This
eventually translates into an instability at large enough $\eta$ \cite{Ducloue:2019ezk}.

\paragraph{Collinear resummations in $\eta$}
As for the evolution
in $Y$ \cite{Beuf:2014uia}, one can avoid this instability via an all-order resummation
leading to an evolution {\it non-local} in $\eta$ \cite{Ducloue:2019ezk}. Unlike what happens with the
resummations in $Y$, the scheme dependence for the resummations in
$\eta$ is small and in agreement with the expected
perturbative accuracy of the resummed equation. Specifically, the 
{\it collinearly-improved} BK equation in $\eta$ reads \cite{Ducloue:2019ezk}
\begin{equation}
	\label{bketa}
	\frac{\del {S}_{\bx\by}(\eta)}{\del \eta}  = 
	\int \dif^2\bz \frac{\abar(r_{\min})}{2\pi} 
	\frac{(\bx-\by)^2}{(\bx-\bz)^2(\bz-\by)^2}\,
	\big[{S}_{\bx\bz}(\eta \minus \delta_{\bx\bz;r})
	{S}_{\bz\by}(\eta \minus \delta_{\bz\by;r}) - {S}_{\bx\by}(\eta) \big], 
\end{equation}
Here, ${S}_{\bx\by}$ is the elastic $S$-matrix  for a dipole with transverse coordinates 
$\bx$ and $\by$, for the quark and the antiquark respectively.
The emission of a soft gluon with transverse position
$\bz$ is viewed as the splitting of the original dipole $(\bx,\,\by)$
into two daughter dipoles $(\bx,\,\bz)$ and $(\bz,\,\by)$.
Besides using the target rapidity $\eta$ (instead of the dipole rapidity $Y$), 
\eqn{bketa} differs from the LO BK equation via the {\it rapidity shift}, defined as
\begin{equation}\label{delta}
  \delta_{\bx\bz;r} \equiv \max \left\{0,\ln\frac{r^2}{|\bx\minus\bz|^2} \right\} 
\end{equation}
with $r\equiv|\bx-\by|$ and similarly for $\delta_{\bz\by;r}$. This
is non-zero only for emissions in which one of the daughter dipoles
is smaller than the parent one. By Taylor-expanding the
rapidity shifts and iteratively using the BK equation for
$\del S/\del\eta$, one generates the whole series of (leading)
double collinear logarithms.
The running coupling corrections are known to be important
and can be minimised by using a one-loop running coupling $\abar(r_{\min})$
with the scale set by the size $r_{\min}$ of the smallest dipole: $r_{\min} \equiv
\min\{|\bx-\by|,|\bx-\bz|,|\bz-\by|\}$.
\eqn{bketa} has to be solved as an initial value problem, with the initial condition
${S}_{\bx\by}(\eta \leq \eta_0) ={S}^{(0)}_{\bx\by}$.
It is possible to extend this equation to full NLO accuracy by adding
the missing pure $\abar^2$ corrections~\cite{Ducloue:2019ezk}.

\begin{figure}
  \centerline{%
    \includegraphics[width=0.35\textwidth,page=1]{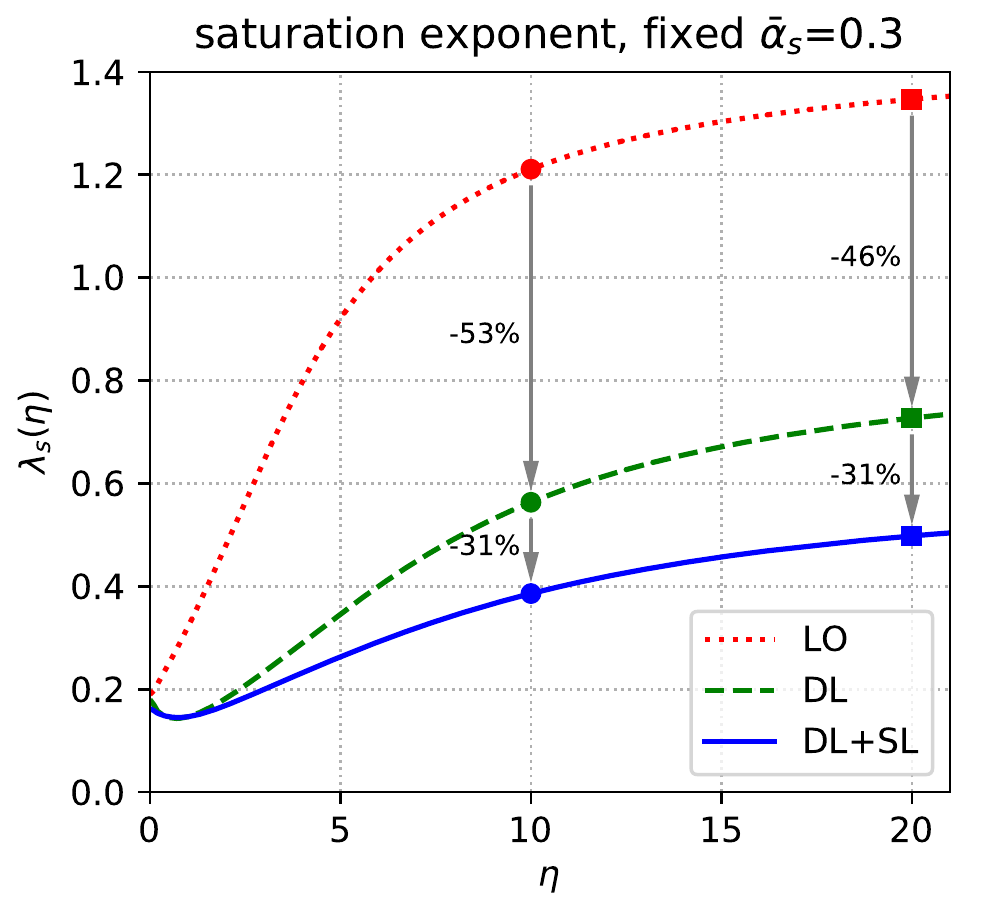}\qquad\qquad%
    \includegraphics[width=0.35\textwidth,page=2]{lambda.pdf}%
  }
  \caption{The saturation exponent $\lambda_s(\eta)$ predicted by
       \eqn{bketa} with Gaussian initial condition  $S_0(r)=\exp[-r^2 Q_0^2]$ with $Q_0^2=1\,{\rm GeV}^2$
and with both
    fixed and running coupling. The lines denoted as DL (green, dashed)  and LO  (red, dotted) are obtained with and
    without the rapidity shift, respectively. The DL+SL (continuous, blue) line corresponds to additionally resumming
    DGLAP-like single logs~\cite{Iancu:2015vea,Ducloue:2019ezk}.}
  \label{fig:lambdas}
\end{figure}

In Fig.~\ref{fig:lambdas} we show the results of \eqn{bketa} 
for the exponent $ \lambda_s(\eta)\equiv {\rmd\ln Q_s^2}/{\rmd\eta}$, with $Q_s(\eta)$ obtained from the
condition that $S(\eta,r)=0.5$ when $r=2/Q_s(\eta)$. 
The effect of the resummation is important, especially for a fixed coupling. When the coupling runs,
the whole evolution slowes down considerably, but the seemingly modest,
additional, reduction ($\sim 20\%$) in $\lambda_s$ due to the
resummation is essential to optimally accommodate the
small-$x$ HERA data. This is demonstrated in
\cite{Ducloue:2019jmy}, where very good fits for the inclusive DIS data 
at $x\le 0.01$ were obtained by using the collinearly improved BK equation
in $\eta$ (with a further inclusion of DGLAP logs, cf.~Fig.~\ref{fig:lambdas}), together with a physically motivated model for the initial condition at $x_0=0.01$.

\paragraph{Acknowledgements} 
Part of the work of B.D, E.I. and G.S. has been supported by the Agence Nationale de la Recherche project ANR-16-CE31-0019-01. Part of the work of B.D has been supported by the ERC Starting Grant 715049 ``QCDforfuture''.












\end{document}